\documentclass{PoS}

\def\to{\rightarrow}
\def\eg{{\it e.g.}}

\title{SUSY Without Prejudice at the 7 and 8 TeV LHC: Gravitino LSPs}

\ShortTitle{SUSY Without Prejudice at the 7 and 8 TeV LHC: Gravitino LSPs}

\author{M.W. Cahill-Rowley, J.L. Hewett, A. Ismail, \speaker{T.G. Rizzo}%
         \thanks{Work supported in part by the Department of Energy, 
Contract DE-AC02-76SF00515}\\
        SLAC National Accelerator Laboratory\\
        E-mail: \email{mrowley,hewett,aismail,rizzo@slac.stanford.edu}}

\abstract{We have examined the capability of the LHC, running at both 7 and 
8 TeV, to explore the 19(20)-dimensional parameter space of the pMSSM with 
neutralino(gravitino) LSPs and soft masses up to 4 TeV employing the ATLAS 
SUSY analysis suite. Here we present some preliminary results for the 
gravitino model set, following the ATLAS analyses whose data were publically 
available as of mid-September 2012. We find that the impact of the reduced 
MET, resulting from models with gravitino LSPs on sparticle searches is more 
than off-set by the detectability of the many possible long-lived NLSPs.}

\FullConference{36th International Conference on High Energy Physics,\\
		July 4-11, 2012\\
		Melbourne, Australia}

\begin{document}

Although searches at the LHC are continuing unabated and a SM-like Higgs 
boson has been recently discovered{\cite {higgs}}, there is still no hint of 
SUSY or any other new physics beyond the Standard Model on the horizon. Several 
conventional SUSY breaking 
scenarios, such as the mSUGRA/cMSSM framework in its most naive form, are now 
highly disfavored by the data. It is clear  
that more model-independent approaches to SUSY which still allow for correlations 
among the many different experimental results are required. We {\cite {us}} have recently 
begun a detailed study of SUSY within the 19-parameter 
p(henomenological)MSSM and have examined the effects of numerous ATLAS-based 
SUSY searches at both 7 and 8 TeV on our model set with neutralino LSPs. This very 
general parametrization allows for a wide range of phenomenological SUSY signatures 
while simultaneously correlating, \eg, collider and dark matter searches.  
Here we will briefly present some of the preliminary results obtained for 
the corresponding 20-parameter  
case with gravitino LSPs. For more details of these analyses and as well as of our overall 
model generation framework and procedures, see{\cite {us}}.

As discussed in previous work{\cite {us}} we reproduce the suite of ATLAS-based 
SUSY searches as closely as possible using a modified version of 
PGS{\cite {PGS}}, employing ATLAS SM backgrounds and validating against 
ATLAS benchmark models. The obvious place to begin such a study is with  
the `vanilla', generalized MET searches at both 7 and 8 TeV since these are powerful 
analyses that usually cover a large fraction of the model parameter space. It is 
important when doing this to include analyses at both center of mass energies,  
since some models which are excluded by the 7 TeV data remain 
allowed at 8 TeV. We previously observed the same effect in our neutralino model 
set, likely as a result of the harder cuts bypassing models with compressed spectra.  
We note that we expect the coverage provided by these MET-based analyses to be somewhat 
diminished overall for the gravitino model set as the many long-lived NLSPs 
occuring at the bottom of decay chains lead to a reduction in the overall amount 
of MET.  However, this is partially compensated by the fact that most of our gravitinos 
are effectively massless leading to large MET in models where the NLSP does decay 
to the gravitino within the inner detector. A further potential compensation is 
provided by the much higher frequency of light binos in the gravitino set which can 
lead in some cases to final states with large MET and additional leptons. Interestingly, 
searches for long-lived sparticles lead to a substantial overall increase in model 
coverage, more than compensating for the reduced MET.  

Tables 1 and 2 show the results of the 7 and 8 TeV `vanilla' MET searches, respectively. 
Here we see several things: ($i$) In comparison to the corresponding results for the 
neutralino models{\cite {us}} the model coverage from jets plus MET is somewhat 
degraded due to reduced MET as we expected 
at both center of mass energies. ($ii$) While the 2-6 jets analysis remains the most powerful 
here, the roles played by the multijet and single lepton analyses at both energies and 
the SSDL analysis at 8 TeV are significantly enhanced with respect to the results 
obtained for neutralino LSPs. Searches with leptons are found to be more effective 
because of lepton production in decays to non-neutralino NLSPs, and also because bino 
NLSPs are common~\cite {us}. ($iii$) Many of the models are found to be excluded by 
multiple analyses. ($iv$) Including the additional requirement on the model set of a 
Higgs boson with a mass of $m_h=126\pm 3$ GeV (here referred to as the ``Higgs Subset'') 
does not significantly alter the amount of 
model coverage although some degradation is observed (as was the case for the 
neutralino set). The magnitude of this degradation is found to be somewhat less for the 
gravitino set. ($v$) We find that $\sim 1.3$k gravitino LSP models are excluded by the 
7 TeV analyses that were {\it not} excluded by the 8 TeV analyses.

\begin{table}
\centering
\begin{tabular}{|l|l|c|c|} \hline\hline
Search  &   Reference          &   Full Model Set  & Higgs Subset \\ \hline
2-6 jets & ATLAS-CONF-2012-033  &  17.76\%  &  16.67\%  \\
multijets & ATLAS-CONF-2012-037  &  2.27\%   & 2.09\%  \\
1-lepton  & ATLAS-CONF-2012-041  &  5.31\%  & 4.63\%  \\
Total  &      &  19.44\%  & 17.95\%  \\
\hline\hline
\end{tabular}
\caption{Fraction of our pMSSM models with gravitino LSPs excluded (in per cent)
 by the general ``vanilla'' MET ATLAS searches at the 7 TeV LHC with 4.7 fb$^{-1}$ of
integrated luminosity for both the full model set as well as for the ``Higgs Sub
set'' satisfying the Higgs mass constraint, $m_h=126\pm 3$ GeV.}
\label{7TeV-vanilla}
\end{table}

\begin{table}
\centering
\begin{tabular}{|l|l|c|c|} \hline\hline
Search  &   Reference          &   Full Model Set  &  Higgs Subset \\ \hline
2-6 jets   &   ATLAS-CONF-2012-109     &  21.83\%  &  20.82\%  \\
multijets   &  ATLAS-CONF-2012-103  &  4.13\%   & 4.02\%  \\
1-lepton     &  ATLAS-CONF-2012-104  &  5.38\%   & 5.05\%  \\
SS dileptons &  ATLAS-CONF-2012-105  & 11.50\%   & 11.14\%  \\
Total        &     & 28.69\%  & 27.19\%  \\
\hline\hline
\end{tabular}
\caption{Same as Table~1 but for the 8 TeV, 5.8 fb$^{-1}$ ATLAS searches.}
\label{8TeV-vanilla}
\end{table}

In Table~3 we see the results for both the heavy flavor(HF) and multilepton(ML) 
ATLAS analyses that probe for light third generation squarks and light gauginos, 
respectively, as might be expected to occur in pMSSM models with low values of 
fine-tuning{\cite {us}}. These searches produce results which are somewhat different 
than those found for neutralino LSPs. Both types of searches, but particularly 
in the case of the ML searches (as would be expected from the above discussion), are  
found to be much more effective for the gravitino models. The somewhat lighter stops 
and sbottoms 
in the gravitino model set also partially enhance the HF search capabilities.

\begin{table}
\centering
\begin{tabular}{|l|l|c|c|} \hline\hline
Search  &   Reference          &   Full Model Set  & Higgs Subset \\ \hline
Gluino $\to$ Stop/Sbottom   &   1207.4686               &  4.06\%   &  4.38\%  \\
Very Light Stop  &    ATLAS-CONF-2012-059               &  0.03\%  &   0.01\%    \\
Medium Stop  &   ATLAS-CONF-2012-071                &  4.92\%   &  4.34\%  \\
Heavy Stop (0l)  &  1208.1447                 &  3.29\%   &  3.87\%  \\
Heavy Stop (1l)   &  1208.2590                &  2.26\%   &  2.51\%  \\
GMSB Direct Stop   &   1204.6736               &  0.05\%  &   0.06\%    \\
Direct Sbottom  &    ATLAS-CONF-2012-106               &  2.80\%   &  2.83\%  \\
3 leptons  &   ATLAS-CONF-2012-108              &  5.91\%   &  5.48\%  \\
1-2 leptons  &    1208.4688               &  8.15\%   &  7.19\%  \\
Direct slepton/gaugino (2l)  &   1208.2884                 &  1.18\%   &  1.02\%  \\
Direct gaugino (3l)  &   1208.3144                &  5.54\%   &  4.82\%  \\
                 &         &       & \\
HF Total    &     &  11.14\%  & 11.30\%    \\
ML Total     &    &  12.10\%   & 10.97\%   \\
\hline\hline
\end{tabular}
\caption{Same as Table~1 but for the HF and ML searches.}
\label{HF-ML}
\end{table}

In Table~4 we see the corresponding results for the non-MET searches. The search for 
heavy stable charged particles(HSCP) is found to be 
significantly more effective in the case of gravitino LSPs due to the high frequency 
of long-lived NLSPs in this set. These gains, plus those for the leptonic searches as 
discussed above, are then responsible for the overall increase in pMSSM model coverage 
that we find for the gravitino pMSSM ($\sim 46\%$) in comparison with the neutralino 
pMSSM ($\sim 34\%$){\cite {us}}. Note that the effect of the Higgs mass cut on the 
{\it total} model space coverage is found to be minimal for both model sets.

\begin{table}
\centering
\begin{tabular}{|l|l|c|c|} \hline\hline
Search     &  Reference   &   Full Model Set  & Higgs Subset \\ \hline
HSCP       &  1205.0272   &  16.93\% &  15.36\%  \\
Dis. Tracks  & ATLAS-CONF-2012-111  &  1.12\%   & 1.16\%  \\
$B_s \to \mu^+\mu^-$  & ATLAS-CONF-2012-061   &  3.11\%   & 6.11\%  \\
$A/H\to \tau^+\tau^-$  &  1202.4083  & 0.07\%   & 0.03\%  \\
     &         &         &        \\
All Searches  & & 45.57\%  & 45.62\%  \\
\hline\hline
\end{tabular}
\caption{Same as Table~1 but now for the non-MET searches. The corresponding 
combined results obtained from all searches is also shown.}
\label{Non-MET}
\end{table}

\begin{figure}[htbp]
\vspace{-5.20cm}
\centerline{\includegraphics[width=0.6\textwidth]{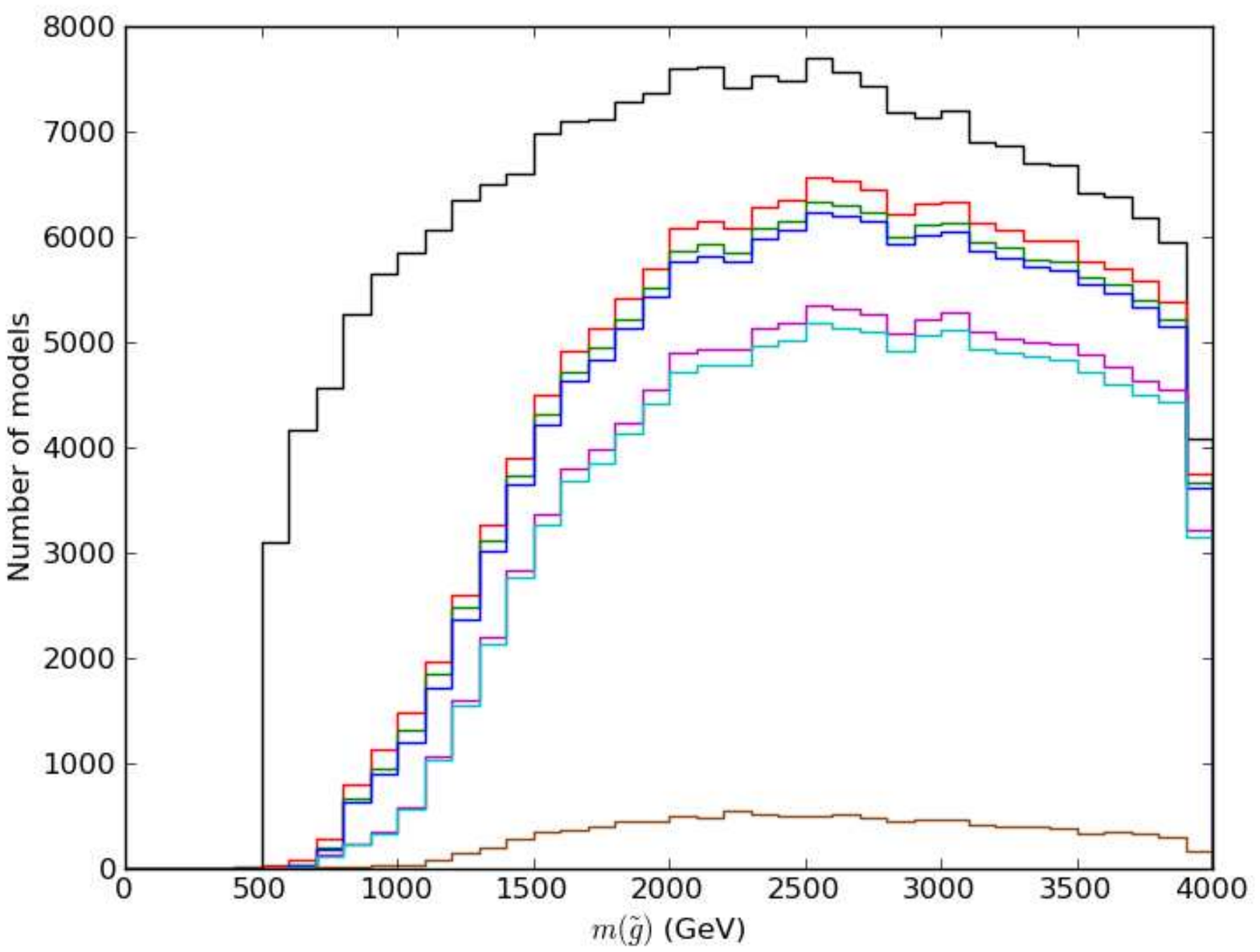}
\hspace{-0.50cm}
\includegraphics[width=0.6\textwidth]{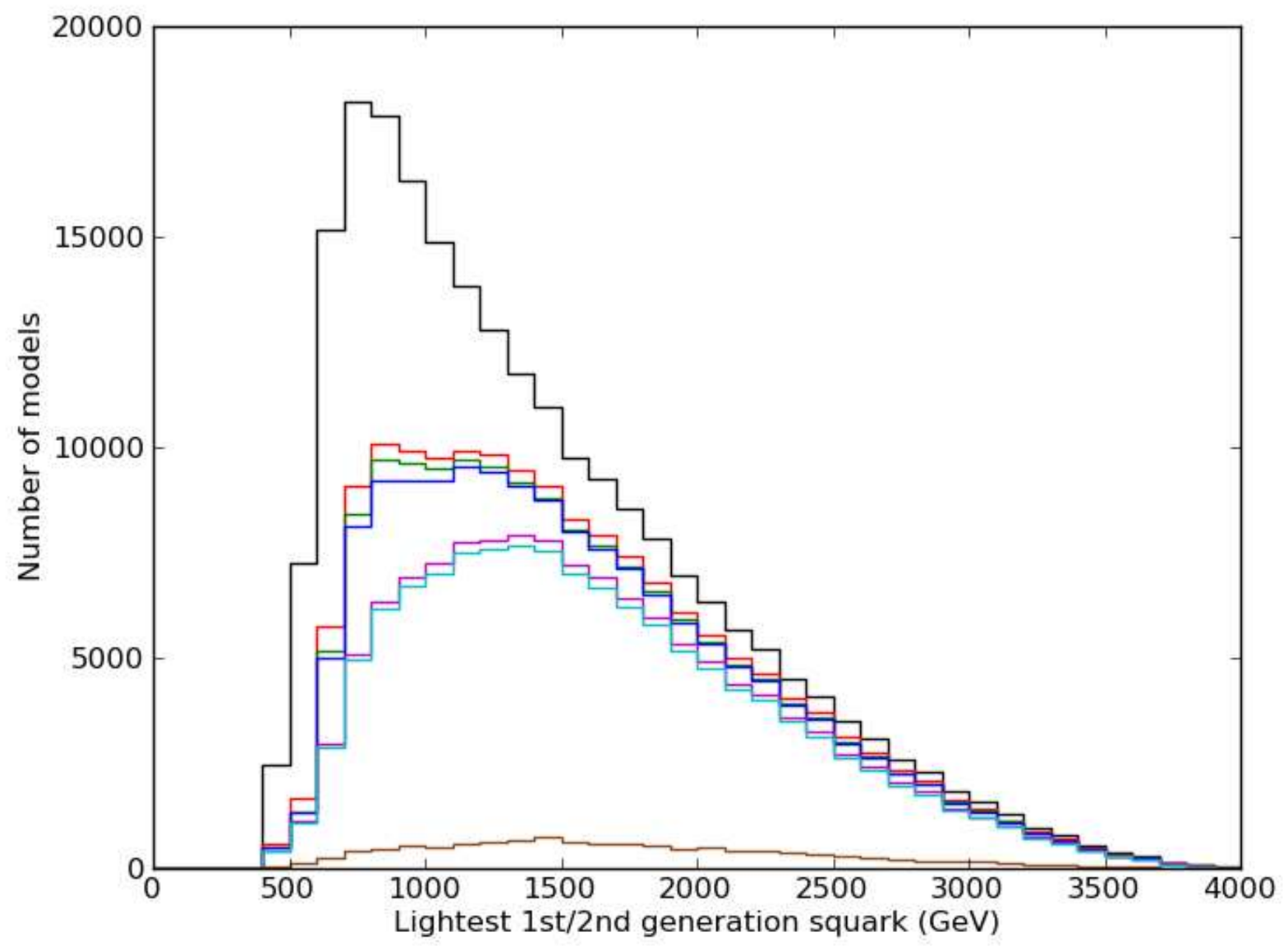}}
\vspace{-5.20cm}
\centerline{
\includegraphics[width=0.6\textwidth]{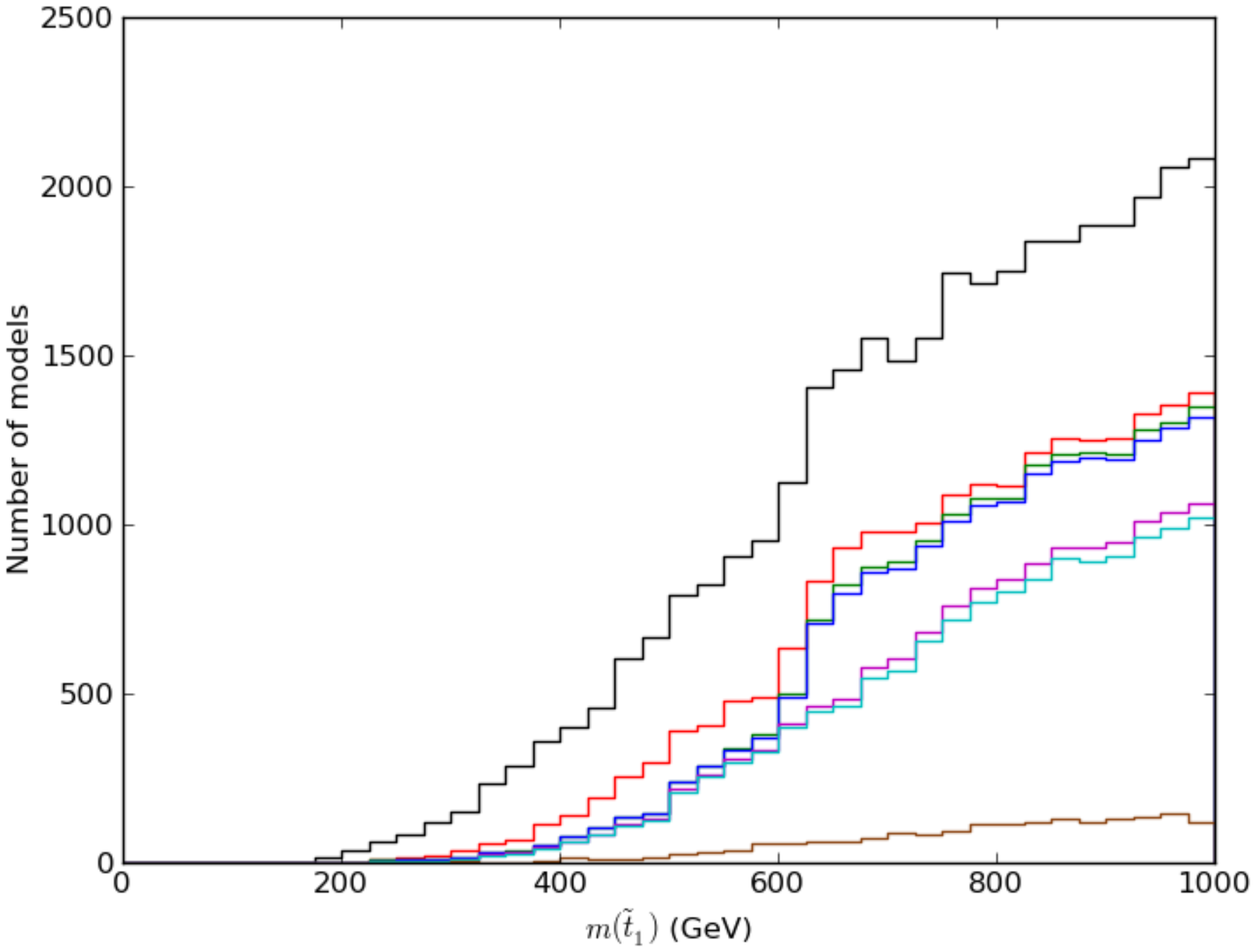}
\hspace{-0.50cm}
\includegraphics[width=0.6\textwidth]{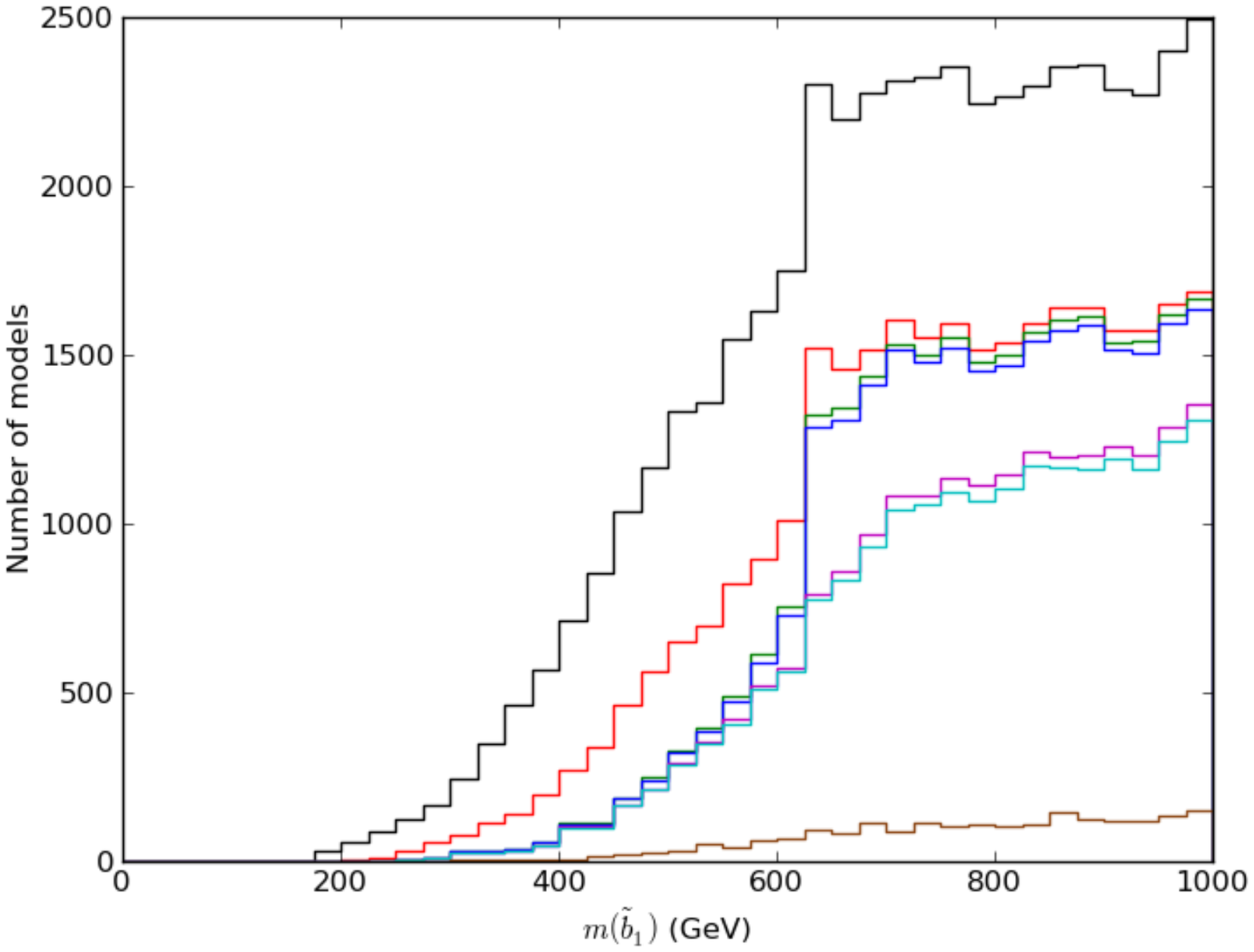}}
\vspace*{-0.10cm}
\caption{Mass distributions for the gluino (top left), lightest first/second 
generation squark (top right), lightest stop (bottom left) and lightest sbottom 
(bottom right) as they respond to the null LHC SUSY searches. The various 
histograms are described in detail in the text.}
\label{fig1}
\end{figure}

How does the gravitino pMSSM parameter space respond to these null searches? 
Fig.~\ref{fig1} presents histograms of the distribution for the gluino, the 
lightest 1st/2nd generation squark, the lightest stop and the lightest sbottom in 
our gravitino model set. The effect of sequentially applying LHC searches on these 
distributions is shown as a series of colored histograms in the following order 
from top to bottom: The original model set as generated (black), 7 and 8 TeV `vanilla' 
searches (red), heavy flavor (green), multileptons (blue), HSCP and disappearing 
tracks (magenta), $B_s\to \mu^+\mu^-$ and $H/A\to \tau^+\tau^-$ (cyan), and 
$m_h=126\pm 3$ GeV (brown).

For both the 
gluinos and the lightest 1st/2nd generation squarks, the vanilla and stable sparticle 
searches are seen to be most effective in excluding models. Overall, these 
distributions are qualitatively similar to those for the neutralino model set. 
Once again, we note the presence of viable models with 
light 1st/2nd generation squarks below 600 GeV, 
gluinos below 700 GeV, and 3rd generation squarks below 400 GeV, although in each 
case the low-mass region is more depleted than in the neutralino model set. (We expect 
some models in the low-mass regions because of the significant number of models with 
a stable neutralino NLSP, producing a scenario identical to the neutralino model set 
except for differences in cosmological constraints.) We also observe that the 3rd 
generation searches remain effective at higher stop and sbottom masses in the gravitino 
set, likely as a result of sensitivity to models in which the 3rd generation squarks 
decay promptly to gravitinos, producing clean signatures with lots of MET.

\begin{figure}[htbp]
\vspace{-5.20cm}
\centerline{\includegraphics[width=0.65\textwidth]{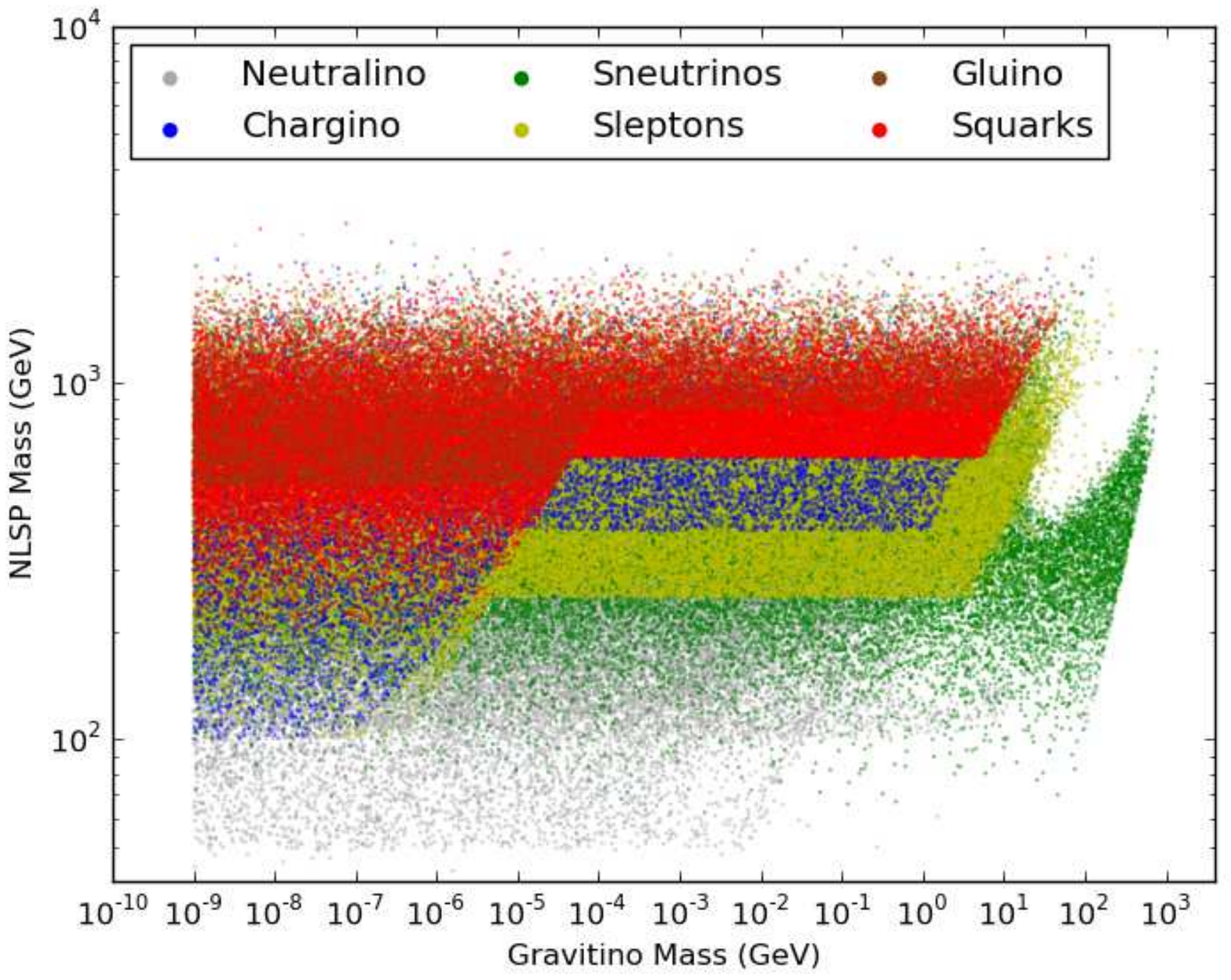}
\hspace{-0.50cm}
\includegraphics[width=0.65\textwidth]{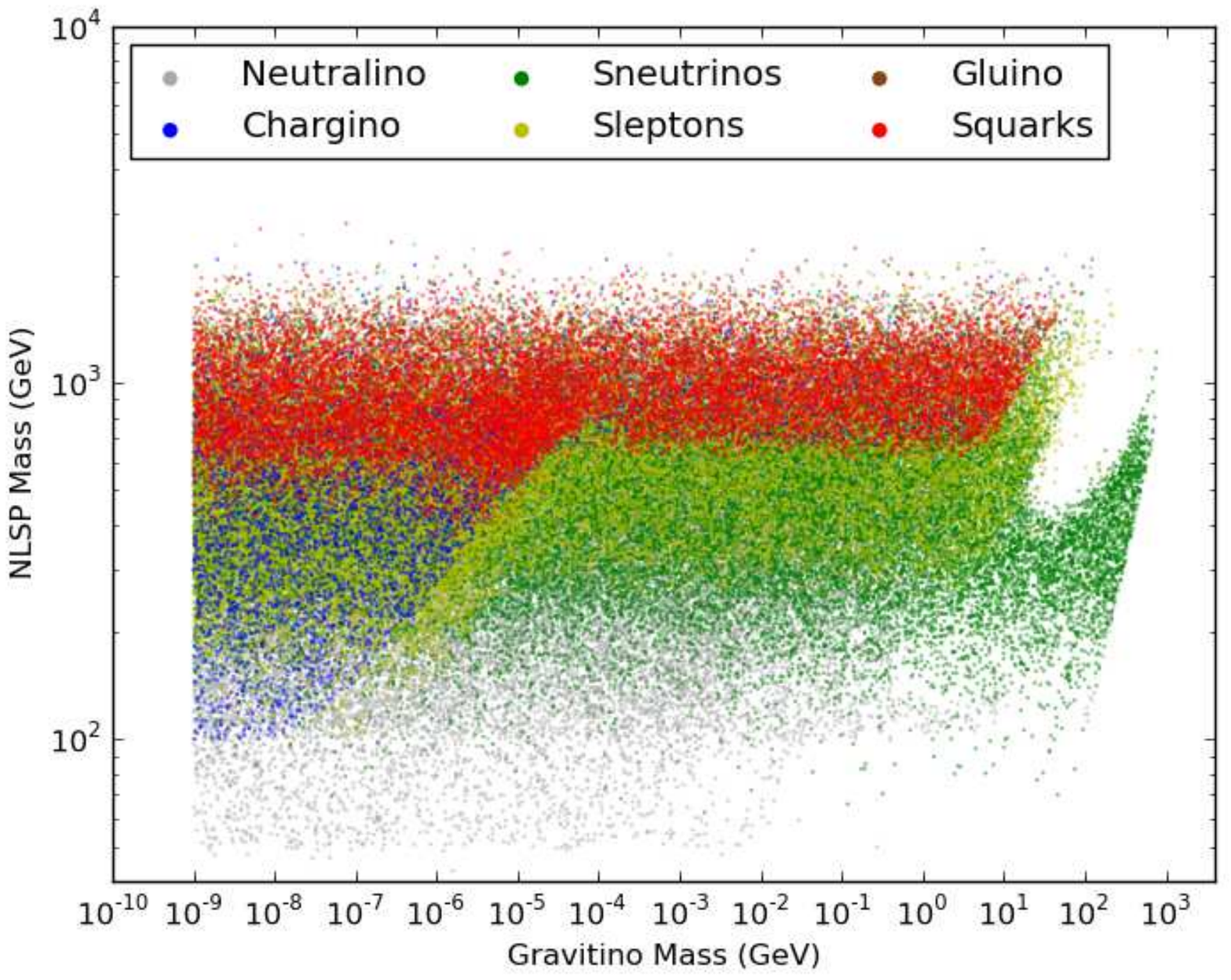}}
\vspace*{-0.10cm}
\caption{Viable models in the NLSP - gravitino mass plane, color-coded according to NLSP
type as labeled. The two plots show the gravitino pMSSM model set before (left) and 
after (right) the LHC constraints (with the exception of the Higgs mass cut) have been 
applied. The horizontal and slanting lines in the plot on the left show the effects of 
model-independent limits on stable particles which have been superseded by a 
model-dependent implementation of LHC HSCP searches. The structure on the far right of 
each graph is produced by the cosmological constraints described in ~\cite {us}.}
\label{fig2}

\end{figure}

In Fig.~\ref{fig2} we display the distribution of masses for the NLSP as a function 
of the gravitino mass for various possible NLSP identities, both as initially generated 
and after applying all of the LHC constraints discussed above except for the Higgs mass 
cut which essentially only reduces the overall statistics. The boundary lines and 
corresponding bands at the model generation level arise due to both the nucleosynthesis 
constraints and those on long-lived sparticles which roughly scale as $\sim M^5/m_{3/2}^2$ 
with $M(m_{3/2})$ being the NLSP(gravitino) mass. The location of the bands themselves 
are correlated with the amount of electromagnetic and/or hadronic energy deposition the 
various NLSP candidates can release via their decay to gravitinos during nucleosynthesis. 
Clearly neutral, color-singlet NLSPs are least constrained by these considerations while,
\eg, colored sparticles are strongly depleted. After 
the various LHC search constraints are applied, we see in the RH panel that some of the 
original bands, particularly those for NLSPs which are charginos, sleptons or squarks in 
the central part of the plot are now essentially absent. In particular, chargino
NLSPs are strongly depleted when the gravitino is heavy enough that the NLSP is stable. 
Stable slepton NLSPs, meanwhile, remain viable at low masses due to small production 
cross-sections, but are uniformly depleted up to high masses ($\tilde 800$ GeV) through 
production in cascade decays and subsequent exclusion by stable particle searches. 
The distribution of models with sneutrino and neutralino NLSPs is 
relatively unaltered, since the exclusion of these models depends on producing charged 
or colored sparticles

In this report we have provided a brief overview of the impact of the ATLAS SUSY 
searches on the pMSSM with gravitino LSPs and contrasted these with the earlier 
results obtained for the neutralino LSP case. Overall we find that due to the 
increased effectiveness of the HF, ML and, most particularly, the long-lived sparticle 
searches, the LHC coverage of the pMSSM model space with gravitino LSPs is substantially   
larger than that for the corresponding neutralino models. For either NLSP choice we 
have found that the restriction to the subset of models with $m_h=126\pm 3$ GeV 
has almost no impact on the fraction of models covered by the complete set of LHC 
searches. Clearly additional searches, particularly ones which are sensitive to long-lived 
sparticles,  will likely be able to further increase the coverage for the gravitino model 
set as we will see in future work~{\cite {us}}.

\end{document}